# Cognitive Green Radio for Energy-Aware Communications

*Malek Naoues, Quentin Bodinier, Jacques Palicot*

*CentraleSupélec/ IETR, {malek.naoues,quentin.bodinier,jacques.palicot}@centralesupelec.fr*

### Introduction

In 5G networks, the number of connected devices, data rate and data volume per area, as well as the variety of QoS requirements, will attain unprecedented scales. The achievement of these goals will rely on new technologies and disruptive changes in network architecture and node design. Energy efficiency is believed to play a key role in complementing the 5G technologies and optimizing their deployment, dynamic configuration and management [1].

Within the framework of green communications and networks, especially for next-generation green cellular radio access networks, the GREAT (Green Cognitive Radio for Energy-Aware wireless communication Technologies evolution) initiative, a ComInLabs Excellence Center (Laboratoire d'Excellence) and Université Européenne de Bretagne (UEB)project, has mainly addressed the fundamental issues of energy efficiency from various perspectives and angles, leveraging on cognitive techniques, at networking level as well as at thephysical layer level.

Various cognitive green radio techniques have been utilized and verified to improve energy efficiency in different application scenarios. Particularly, representative learning algorithms and decision-making schemes (transfer learning, reinforcement learning, combined learning, entropy theory, etc.) [2] have been successfully employed to achieve energy saving under spectral/capacity requirements.

In order to improve the balance of energy-efficiency and spectrum-efficiency, the fundamental characteristics of networking traffics have been investigated by taking into account the mobile users' behavior (mobility, location/position/density, social connection, living style, etc.), in various outdoor as well as indoor networking environments[3]. The various led theoretical analysis have been validated through implementation in the real world. As a case in point, we present in this work a Wi-Fi-based platform which implements two main functionalities: first, users clustering and second, base station (BS) & access points (AP) switchingon/off [4].

This paper is organized as follows: in the first section the concept of cognitivegreen radio is presented. In the second section we introduce our RSS-based clustering approach. The third section focuses on topology management under energy consumption constraint. The fourth section presents how these two different concepts were implemented in our Wi-Fi-based platform.

### 1. The Cognitive Green Radio

Cognitive Green Radio (CGR) is a Cognitive Radio which is aware of sustainable development and takes it as an additional constraint in the decision making function of the cognitive cycle [1]. The implementation of these new approaches and technologies will rely on advanced PHY/MAC techniques calling for a variety of coordination and cooperation strategies among the network elements. Furthermore, it is envisioned that future networks will be complemented by cognitive entities that exploit context information to optimizetheir performances and their energy-efficiency. Such entities should be based on the cognitive cycle summarized in three key steps according to [1]:

- **Observe (Sense):** It makes the context information available to the system. It relies on various kinds of available sensing means from the operating environment.
- **Decide (Learn, Plan and Decide):** It implies various kinds of intelligence, including learning, planning, and decision making.
- **Act and Adapt:** It corresponds to the dynamic configuration and management of the various network elements.

We emphasize here that the space of context information is not restricted to spectrum availability, but can include any kind of information relevant to the decision making and adaptation steps of the cognitive cycle. One example of context information is location information, i.e. the locations of the users and connected devices.

In this paper, we propose the received signal strength (RSS) of connected devices as a kind of context information. Here a connected device refers to user equipment (UE) or any other kind of portable or non-portable device with wireless connectivity (e.g., smart grid components, metering units, sensors, etc.).

## 2. RSS-Based Clustering

Clustering is the task of partitioning a data set into several groups, called clusters, in such a way that the elements in the same cluster are more similar to each other than to those in other clusters. Thedetection of clusters adds intelligence to the network, and canenhance some energy efficient schemes at network level, suchas topology management, relaying, beamforming, offloading,and heterogeneous networks deployment.

In the clustering problem considered here the spatial proximity of devices will reflect their similarity. One approach to find the clusters is to rely on the devices locations, i.e., their coordinates in a 2D or 3D Euclidean space. However, finding the locations for a large a number of devices, say hundreds or thousands, could incur high communication and computation costs, and the performance will depend on the location estimation accuracy. RSS-based clustering is enabled by the fact that two nearby devices have a high probability of measuring close RSS values with one BS. The clustering operation starts by collecting the RSS measurements made with the devices to a central processor, and associating an RSS vector to every device, where an entry of the vector is obtained with one BS or network node. Fig. 1(a) shows an example of a matrix of RSS values collected in a Wi-Fi network [4]. Each row of this matrix represents the RSS vector for one device. Afterwards, a clustering algorithm is applied to the RSS vectors. The algorithm needs to solve the following problems:

- Identifying the 'clutter' which is the set of isolated devices that do not have other devices in their nearby vicinity, as illustrated in Figure 1.
- Finding the clusters for the remaining non-clutter devices. The number of clusters is not known in advance and also needs to be computed.

The authors in [5] developed a non-parametric solution for clustering algorithm. The application of this solution to the RSS matrix of Fig. 1(a) yields the results shown in Fig. 1(b).

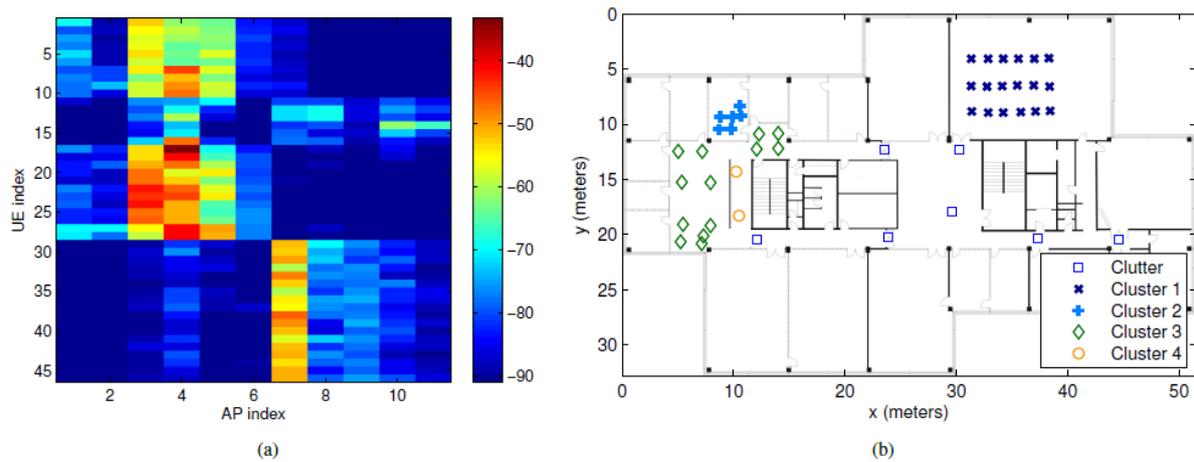

*Fig.1 (a) RSS matrix (in dBm) for 46 UEs obtained with 11 Wi-Fi access points. The UEs locations are plotted in Fig. 1(b). (b) Clutter detection and clustering result obtained by applying the non-parametric algorithm developed in [5] to the RSS matrix of Fig. 1(a).*

Fig.2 shows an example of the non-parametric clustering for 1-dimensional data points. We call 'data point' the RSS vector associated with a device. The non-parametric clustering assumes that the data points are randomly and independently generated according to a distribution function $f$, and the high density points belonging to clusters are those verifying $f(\text{data point}) > c$, where c is a given constant. The remaining data points correspond to the clutter. Since $f$ is unknown, it is estimated using a non-parametric kernel-based method [5]. $\hat{f}$ denotes the estimate of the distribution function $f$ in Fig. 2.

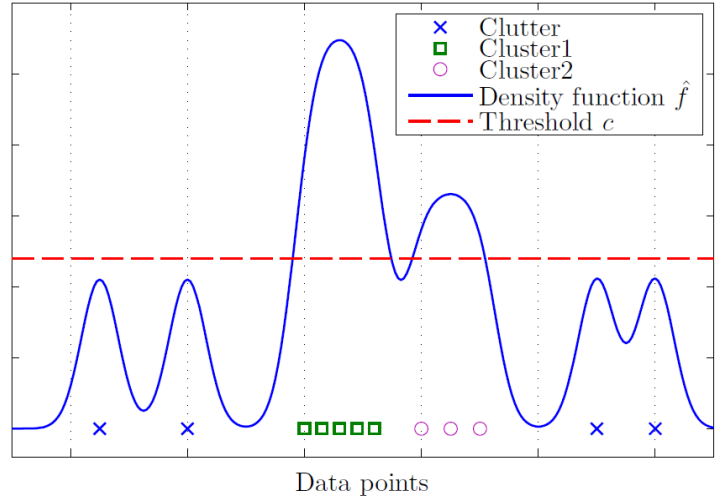

*Fig.2 Non-parametric clutter detection and clustering. The function $\hat{f}$ is obtained by placing a Gaussian kernel at every data point. The value of $\hat{f}$ for the clutter is below the threshold. There are 2 clusters and 2 modes above the threshold [5].*

A measurement campaign was conducted to evaluate the clustering performance in an indoor environment based on Wi-Fi technology. RSS measurements were collected by a PC placed on a moving trolley over a grid of points covering a big portion of the area of Figure 1(b). The rooms are separated by wooden and concrete walls, and the access points (APs) are placed on multiple floor levels. The simultaneous presence of several UEs is emulated by randomly drawing several points from the grid. To generate a cluster, a space zone is randomly selected, and then points are uniformly drawn from the gridpoints belonging to this zone. The clutter UEs are uniformly drawn after removing the grid points lying inside the selected cluster zones. The performance of classifying the UEs as clutter or belonging to clusters can be measured in terms of the probability of false alarm (PFA) which is the probability of wrongly classifying a clutter UE as belonging to a cluster, and the probability of detection (PD) which is the probability of correctly classifying a cluster UE. These probabilities are computed by averaging the outcomes of 100 trials, and are plotted in Figure 3. Figure 3(a) shows that PFA is increasing with the clutter density. When the number of used APs is high (i.e., 6 and 11 APs), the variation of PFA with the number of clusters was not much noticeable. This might be due to the fact that the RSS vectors have a higher dimension for a higher number of APs, making the clutter RSS vectors farther from the clusters ones. It was not noticed that PD varies with the clutter density, however it varies with the number of APs, and cluster size and density, as shown in Figure 3(a). Figure 3 reveals a good performance for a clutter density below 0.02UEs/m$^2$ and a cluster density of the order of 0.2 UEs/m$^2$.

For outdoor environments, a good performance based on simulations was obtained in [5] with appropriate clutter and cluster densities. Outdoor environments mainly differ from the indoor ones in that the shadowing correlation function decreases more slowly with the separating distance between the devices.

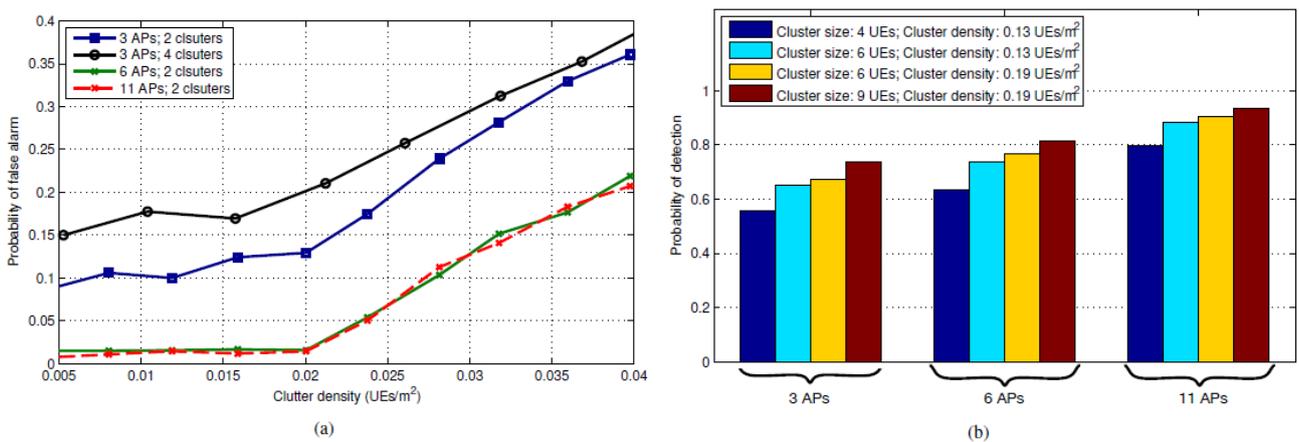

*Fig.3 Performance of the classification of UEs as clutter or belonging to clusters for the indoor Wi-Fi experimentation. (a) Variation of PFA with the clutter density. (b) Variation of PD with the number of APs, cluster size and cluster density.*

## 3. Energy-Aware Network Topology Management

Cellular networks are typically deployed in such a way that a minimum level of QoS is guaranteed during the peak traffic-load periods. However, the traffic loads at a BS vary widely over time. Topology management (or cell shaping) techniques aim at minimizing the energy consumption by turning off some BSs when their traffic load is low, and serving their connected users by the neighboring active BSs. These techniques have been proposed for application to Wi-Fi, small cell and cellular networks.

The detection of the clusters could be useful for these algorithms. For example, as illustrated in Fig.4, favoring the switching off for a BS without clusters since its users are likely to be far from each other and get served by several nearby BSs, and avoiding this operation for a BS with a large cluster to avoid the handover of a large number of users to a single neighboring BS. The appearance of clusters far from active BSs can be also used as a criterion for activating some BSs.

In this work, we recast our topology management problem as an optimization problem, where the object is to maximize the number of APs to be switched off subject to constraints on coverage and QoS. The transmit power of the APs are among the variables of the optimization problem. This problem is then solved using heuristic techniques. The real traffic requirements of mobile users can be obtained by monitoring their activity over a time window. The main constraints of the optimization problem are derived from the RSS measurements made between these UEs and the APs as shown in the matrix of Fig. 1(a). Each row corresponds to one UE and each column corresponds to one AP. An RSS entry in this matrix is obtained by averaging several RSS measurements over a short time window and after eliminating outlying values. The outlier elimination is done automatically by applying a model selection technique [4]. A QoS measured in terms of the data rate should be guaranteed for all the users. The real traffic requirements of users can be obtained by monitoring their activity.

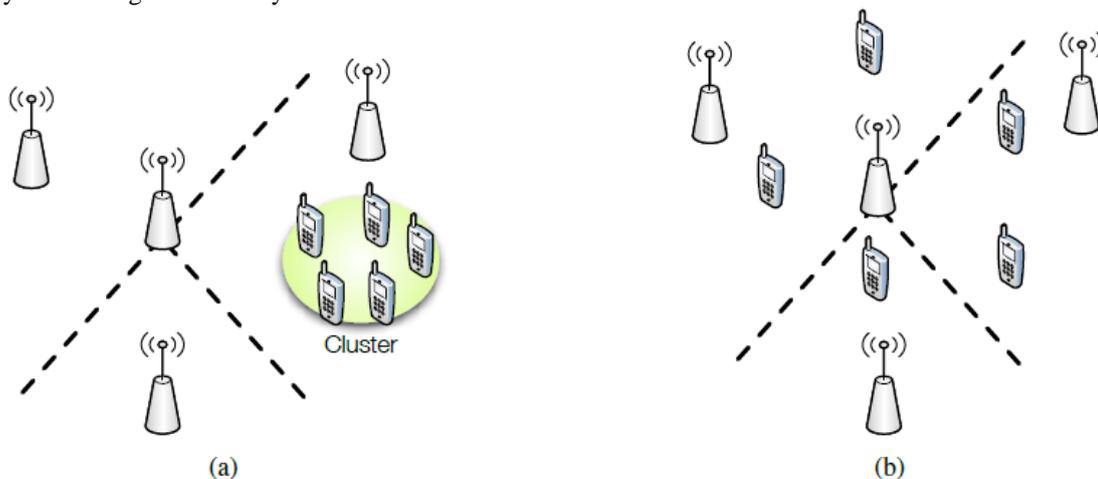

*Fig.4 (a) Cluster of mobile terminals served by the central BS. When this BS is turned off, the cluster of users will make a handover to only one BS (the upper right one) whose traffic might highly increase. (b) There is no cluster in the area of the central BS. When this BS is turned off, its users will get served by several nearby BSs and their traffic will be distributed among them.*

4. **Real-World Implementation on a Wi-Fi based Platform**

In order to validate the various proposed techniques, we developed a Wi-Fi based platform which implements the two aforementioned applications. The main reason behind using a Wi-Fi-based platform to emulate a cellular network is to avoid using proprietary hardware systems and to use unlicensed Industrial, Scientific and Medical (ISM) radio bands. The proposed platform (Fig. 5) is based on several managed 802.11g APs that can be remotely controlled by a server via a WLAN Switch (Switch On/Off, control their transmit power, etc.). The APs are powered via Ethernet and their power consumption is monitored at the WLAN Switch interface. The UEs used in this demonstration are Android smartphones and Linux laptops. These UEs periodically monitor the network and send information about the neighboring APs like RSS measurements, delays and the bandwidth. This information is transmitted via Wi-Fi to the server. The server collects information about the UEs connectivity, handover, etc. The tested algorithms are implemented on the server by running Matlab and Python scripts.

A demonstration of the running platform is available online [6]. This platform has been successfully used to perform user clustering and turn off base stations of the network when they are not needed. This proof of concept showed that Cognitive Green Radio is a paradigm that can be applied in the real world given that enough information is known by the network.

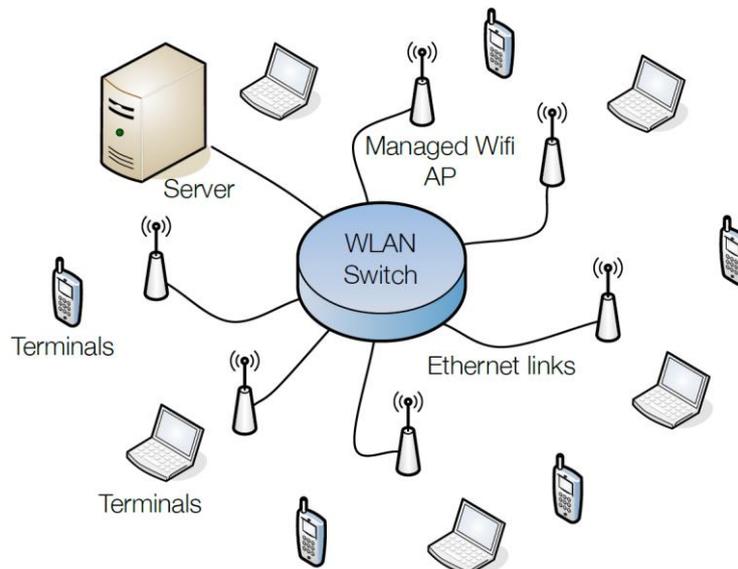

*Fig.5 The proposed platform based on managed 802.11g APs remotely controlled by a server via a WLAN Switch*

## 5. Conclusion

We present in this work some RSS based approaches to achieve energy saving for WLAN and cellular networks. The proposed techniques are tested in real radio environment with two application scenarios: energy-aware topology management for cellular networks and mobile terminals clustering. A real world Wi-Fi based proof of concept is described, which shows that Cognitive Green Radio, far from being a purely theoretical concept, is a realistic, feasible paradigm upon which future wireless networks should be designed.

## Citations

1- J. Palicot, "Cognitive radio: an enabling technology for the green radio communications concept," In Proceedings of the 2009 International Conference on Wireless Communications and Mobile Computing: Connecting the World Wirelessly (IWCMC '09).
2- R. Li, Z. Zhao, X. Chen, J.Palicot, and H. Zhang, "TACT: A Transfer Actor-Critic Learning Framework for Energy Saving in Cellular Radio Access Networks," IEEE Trans. on Wireless Communications, April 2014.
3- R. Li, Z. Zhao, X. Zhou, J.Palicot, H. Zhang, "The Predictability Analysis of Cellular Networks Traffic: From Entropy Theory to Networking Practice," IEEE Communications Magazine, June 2014.
4- M. Naoues, H. Noureddine, Q. Bodinier, H. Zhang, J. Palicot, "Wi-Fi-Based Platform for Energy Saving in Wireless Networks," IEEE Online GreenComm 2014.
5- H. Noureddine, H. Zhang, and J. Palicot, "RSS-Based Clustering of Mobile Terminals for Localization in Wireless Networks," in Proc. Int. Symp. on Wireless Communication Systems, ser. ISWCS2014, Aug.2014.
6- SCEE Research Team, "IEEE Online GreenComm 2014 - Wi-Fi-Based Platform for Energy Saving in Wireless Networks" [YouTube video]. Available: https://youtu.be/Cg3h5vPBv9g Accessed Feb. 2016.